# Leaky Modes of Dielectric Cavities

Masud Mansuripur[†], Miroslav Kolesik[†], and Per Jakobsen[‡]

[†]College of Optical Sciences, The University of Arizona, Tucson
[‡]Department of Mathematics and Statistics, University of Tromsø, Norway

**Abstract**. In the absence of external excitation, light trapped within a dielectric medium generally decays by leaking out — and also by getting absorbed within the medium. We analyze the leaky modes of a parallel-plate slab, a solid glass sphere, and a solid glass cylinder, by examining those solutions of Maxwell's equations (for dispersive as well as non-dispersive media) which admit of a complex-valued oscillation frequency. Under certain circumstances, these leaky modes constitute a complete set into which an arbitrary distribution of the electromagnetic field residing inside a dielectric body can be expanded. We provide completeness proofs, and also present results of numerical calculations that illustrate the relationship between the leaky modes and the resonances of dielectric cavities formed by a simple parallel-plate slab, a glass sphere, and a glass cylinder.

**1. Introduction**. A parallel plate dielectric slab, a solid glass sphere, and a solid glass cylinder are examples of material bodies which, when continually illuminated, accept and accommodate some of the incident light, eventually reaching a steady state where the rate of the incoming light equals that of the outgoing. By properly tuning the frequency of the incident light, one can excite resonances, arriving at conditions under which the optical intensity inside the dielectric host exceeds, often by a large factor, that of the incident light beam. If now the incident beam is suddenly terminated, the light trapped within the host begins to leak out, and, eventually, that portion of the electromagnetic (EM) energy which is not absorbed by the host, returns to the surrounding environment.

The so-called leaky modes of a dielectric body are characterized by their unique complex-valued frequency $\omega_q = \omega_q' + i\omega_q''$, where the index $q$ identifies individual modes [1-5]. The imaginary part $\omega_q''$ of each such frequency signifies the decay rate of the leaky mode. In the following sections, we analyze the EM structure of the leaky modes of dielectric slabs, spheres, and cylinders, and examine the conditions under which an initial field distribution can be decomposed into a superposition of leaky modes. We also present numerical results where the resonance conditions and quality factors ($Q$-factor $= |\omega_q'|/|\omega_q''|$) of certain cavities are computed; the correspondence between these and the leaky-mode frequencies is subsequently explored.

The present paper's contribution to the mathematics of open systems is a completeness proof for leaky modes of dispersive media under certain special circumstances. These include the cases of (i) a dielectric slab initially illuminated at normal incidence, (ii) a solid dielectric sphere under arbitrary illumination, and (iii) a solid dielectric cylinder illuminated perpendicular to the cylinder axis. Our completeness proof, while relatively simple, is self-contained in the sense that it does not rely on any general theorems as is the case, for instance, with quantum mechanical proofs of completeness that rely on the completeness of the Hamiltonian eigenstates.

We begin by analyzing in Sec.2 a non-dispersive dielectric slab illuminated at normal incidence. The analysis is then extended in Sec.3 to the case of a dispersive slab, where we introduce a general methodology for proving the completeness of leaky modes under special circumstances. Numerical results showing the connection between the resonances of the slab (when illuminated by a tunable source) and the leaky mode frequencies are presented in Sec.4. The next section describes the leaky modes of a dielectric slab illuminated at oblique incidence.

Here we find that, although leaky modes exist and can be readily evaluated by numerical means, the proof of completeness encounters a roadblock due to certain mathematical difficulties.

In Sec.6 we discuss the leaky modes of a dispersive dielectric sphere, and demonstrate their completeness for a general class of initial conditions. Numerical results that show the circumstances under which a solid glass sphere resonates with an incident EM field, and also the correspondence between the resonance lines and the leaky-mode frequencies are the subjects of Sec.7. We proceed to extend our methodology to dispersive dielectric cylinders in Sec.8, where we derive the characteristic equation for leaky modes under general circumstances, and provide a completeness proof for these leaky modes in certain special cases where the direction of illumination is perpendicular to the cylinder axis. Numerical results that show strong similarities between the resonances of dielectric cylinders and those of dielectric spheres are presented in Sec.9. The final section provides a summary of the paper followed by concluding remarks.

**2. Leaky modes of a parallel-plate dielectric slab**. Figure 1 shows a transparent slab of thickness $d$ and refractive index $n$, placed in contact with a perfect reflector. At first, we assume the dielectric is free from dispersion, so that, across a broad range of frequencies, $n$ remains constant. (The analysis will be extended in the following section to cover dispersive media as well.) Inside the slab, a standing wave is initially set up by a normally-incident plane-wave (not shown), which oscillates at a frequency $\omega_0$ and is linearly-polarized along the $x$-axis, having counter-propagating $E$-field amplitudes $\pm E_0 \hat{x}$. The incident beam is abruptly terminated at $t = 0$, causing the field inside the slab to leak out and, eventually, to vanish.

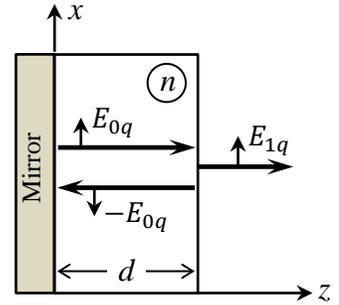

**Fig.1**. Dielectric slab of thickness $d$ and refractive index $n$, coated on its left facet with a perfect reflector. Also shown is a leaky mode, having amplitude $E_{0q}$ inside the slab and $E_{1q}$ outside.

Each leaky mode of this simple cavity may be described in terms of plane-waves having complex frequencies $\omega_q = \omega_q' + i\omega_q''$ and corresponding $k$-vectors $\pm(n\omega_q/c)\hat{z}$ inside the dielectric slab, and $(\omega_q/c)\hat{z}$ outside. Denoting by $\pm E_{0q}\hat{x}$ the amplitudes of counter-propagating plane-waves inside the slab, and by $E_{1q}\hat{x}$ the amplitude of the leaked plane-wave in the free-space region outside, we write expressions for the $E$ and $H$ field distributions of each leaky mode, and proceed to obtain the leaky mode frequencies by matching the boundary conditions at the exit facet of the cavity located at $z = d$.

When the incident beam is abruptly terminated at $t = 0$, the EM field at $t > 0$ may be described as a superposition of leaky modes. Each leaky mode consists of two counter-propagating plane-waves inside the dielectric slab, and a third plane-wave propagating in the free-space region outside. Considering that the $E$-field at the surface of the perfect conductor (located at $z = 0$) must vanish, the $E$ and $H$ fields inside and outside the slab may be written as

$$\boldsymbol{E}_{\text{in}}(\boldsymbol{r}, t) = E_{0q}\hat{\boldsymbol{x}} \exp[i\omega_q(nz - ct)/c] - E_{0q}\hat{\boldsymbol{x}} \exp[-i\omega_q(nz + ct)/c], \tag{1a}$$

$$\boldsymbol{H}_{\text{in}}(\boldsymbol{r}, t) = (nE_{0q}/Z_0)\hat{\boldsymbol{y}} \exp[i\omega_q(nz - ct)/c] + (nE_{0q}/Z_0)\hat{\boldsymbol{y}} \exp[-i\omega_q(nz + ct)/c]. \tag{1b}$$

$$\boldsymbol{E}_{\text{out}}(\boldsymbol{r}, t) = E_{1q}\hat{\boldsymbol{x}} \exp[i\omega_q(z - ct)/c], \tag{2a}$$

$$\boldsymbol{H}_{\text{out}}(\boldsymbol{r}, t) = (E_{1q}/Z_0)\hat{\boldsymbol{y}} \exp[i\omega_q(z - ct)/c]. \tag{2b}$$

Here $Z_0 = \sqrt{\mu_0/\varepsilon_0}$ is the impedance of free space. It is not difficult to verify that each of the plane-waves appearing in the above equations satisfies Maxwell's equations for real- as well as complex-valued $\omega_q$. In general, the refractive index $n$ of the dielectric material should be a function of $\omega_q$, although in the present section we are going to assume that the material is free from dispersion. Note that the two plane-waves inside the slab have equal magnitudes but a relative phase of $\pi$, so that the net $E$-field at the mirror surface (located at $z = 0$) vanishes. The boundary conditions at the exit facet of the dielectric slab (located at $z = d$) now yield

$$E_{0q} \exp(in\omega_q d/c) - E_{0q} \exp(-in\omega_q d/c) = E_{1q} \exp(i\omega_q d/c), \tag{3a}$$

$$nE_{0q} \exp(in\omega_q d/c) + nE_{0q} \exp(-in\omega_q d/c) = E_{1q} \exp(i\omega_q d/c). \tag{3b}$$

Dividing Eq.(3a) by Eq.(3b) eliminates both $E_{0q}$ and $E_{1q}$, yielding the following constraint on acceptable values of $\omega_q$:

$$\exp(i2n\omega_q d/c) = -(n+1)/(n-1). \tag{4}$$

Assuming that $n > 1$, it is clear from Eq.(4) that the imaginary part of $\omega_q$ must be negative. Acceptable values of $\omega_q = \omega_q' + i\omega_q''$ may now be found from Eq.(4), as follows:

$$\omega_q' = \frac{(2q+1)\pi c}{2nd} \quad \text{and} \quad \omega_q'' = -\left(\frac{c}{2nd}\right)\ln\left(\frac{n+1}{n-1}\right). \tag{5}$$

The index $q$ appearing in the above expression for $\omega_q'$ is an arbitrary integer (zero, positive, or negative), which uniquely identifies individual modes of the leaky cavity. Note that, in the absence of dispersion, the imaginary part of $\omega_q$ is independent of the mode number $q$; as such we shall henceforth remove the subscript $q$ from $\omega_q''$, and proceed to write it simply as $\omega''$. Thus the various modes are seen to have different oscillation frequencies, $\omega_q'$, but amplitudes that obey the same temporal decay factor, $\exp(\omega'' t)$.

The beam that leaks out of the cavity and into the free-space region $z > d$, is seen to *grow* exponentially along the $z$-axis, in accordance with the expression $E_{1q} \exp[-\omega''(z-ct)/c]$, but of course this exponential growth terminates at $z = ct$, where the leaked beam meets up with the tail end of the beam that was originally reflected from the front facet of the device (i.e., prior to the abrupt termination of the incident beam at $t = 0$). The EM energy in the region $d < z < ct$ is just the energy that has leaked out of the dielectric slab, with the exponential decrease of the amplitude in time compensating the expansion of the region "illuminated" by the leaked beam.

Inside the dielectric slab, where $0 \le z \le d$, the individual mode profiles of the EM field may be written as

$$\boldsymbol{E}(\boldsymbol{r},t) = E_{0q}\hat{\boldsymbol{x}} \{\exp[-(\omega'' - i\omega_q')nz/c] - \exp[(\omega'' - i\omega_q')nz/c]\} \exp[(\omega'' - i\omega_q')t], \tag{6a}$$

$$\boldsymbol{H}(\boldsymbol{r},t) = (nE_{0q}/Z_0)\hat{\boldsymbol{y}} \{\exp[-(\omega'' - i\omega_q')nz/c] + \exp[(\omega'' - i\omega_q')nz/c]\} \exp[(\omega'' - i\omega_q')t]. \tag{6b}$$

The EM field residing in the dielectric slab at $t = 0$ may be expressed as a superposition of the leaky modes of Eq.(6), so that each mode would evolve in time, oscillating in accordance with its own phase-factor $\exp(-i\omega_q' t)$, while declining in magnitude in accordance with the (common) amplitude-decay-factor $\exp(\omega'' t)$. Unfolding the modal field profile of Eq.(6) around the $z = 0$ plane, then writing the unfolded field at $t = 0$ as a distribution over the interval $-d \le z \le d$, we will have

$$\boldsymbol{E}(\boldsymbol{r}, t=0) = E_{0q}\hat{\boldsymbol{x}} \exp[-(\omega'' - i\omega'_q)nz/c], \tag{7a}$$

$$\boldsymbol{H}(\boldsymbol{r}, t=0) = (nE_{0q}/Z_0)\hat{\boldsymbol{y}} \exp[-(\omega'' - i\omega'_q)nz/c]. \tag{7b}$$

Substitution for $\omega'_q$ and $\omega''$ from Eq.(5) into Eq.(7) now yields

$$\boldsymbol{E}(\boldsymbol{r}, t=0) = E_{0q}\hat{\boldsymbol{x}} \, [(n+1)/(n-1)]^{z/2d} \exp(i\pi z/2d) \exp(iq\pi z/d), \tag{8a}$$

$$\boldsymbol{H}(\boldsymbol{r}, t=0) = (nE_{0q}/Z_0)\hat{\boldsymbol{y}} \, [(n+1)/(n-1)]^{z/2d} \exp(i\pi z/2d) \exp(iq\pi z/d). \tag{8b}$$

It is thus clear that any initial field distribution inside the dielectric slab can be unfolded around the $z = 0$ plane, multiplied by $[(n+1)/(n-1)]^{-z/2d} \exp(-i\pi z/2d)$, then expanded in a Fourier series to create a superposition of the leaky modes given by Eq.(6).

**3. Effects of dispersion**. Suppose now that the dielectric material is dispersive. The simplest case would involve a medium whose electric and magnetic dipoles behave as single Lorentz oscillators, each having its own resonance frequency $\omega_r$, plasma frequency $\omega_p$, and damping coefficient $\gamma$. The electric and magnetic susceptibilities of the material will then be given by

$$\chi_e(\omega) = \frac{\omega_{pe}^2}{\omega_{re}^2 - \omega^2 - i\gamma_e\omega}, \qquad \chi_m(\omega) = \frac{\omega_{pm}^2}{\omega_{rm}^2 - \omega^2 - i\gamma_m\omega}. \tag{9}$$

The corresponding refractive index, now a function of the frequency $\omega$, will be

$$n(\omega) = \sqrt{\mu\varepsilon} = \sqrt{(1+\chi_m)(1+\chi_e)} = \sqrt{1 + \frac{\omega_{pm}^2}{\omega_{rm}^2 - \omega^2 - i\gamma_m\omega}} \times \sqrt{1 + \frac{\omega_{pe}^2}{\omega_{re}^2 - \omega^2 - i\gamma_e\omega}}$$

$$= \sqrt{\frac{(\omega - \Omega_{1m})(\omega - \Omega_{2m})}{(\omega - \Omega_{3m})(\omega - \Omega_{4m})}} \times \sqrt{\frac{(\omega - \Omega_{1e})(\omega - \Omega_{2e})}{(\omega - \Omega_{3e})(\omega - \Omega_{4e})}}, \tag{10a}$$

where

$$\Omega_{1,2} = \pm\sqrt{\omega_r^2 + \omega_p^2 - \tfrac{1}{4}\gamma^2} - \tfrac{1}{2}i\gamma, \tag{10b}$$

$$\Omega_{3,4} = \pm\sqrt{\omega_r^2 - \tfrac{1}{4}\gamma^2} - \tfrac{1}{2}i\gamma. \tag{10c}$$

Assuming that $\gamma \ll \omega_r$, the poles and zeros of both $\mu(\omega)$ and $\varepsilon(\omega)$ will be located in the lower-half of the complex $\omega$-plane, as shown in Fig.2. The dashed line-segments in the figure represent branch-cuts that are needed to uniquely specify each square-root function appearing on the right-hand side of Eq.(10a). For the sake of simplicity, we shall further assume that the branch-cuts of $\sqrt{\mu}$ and those of $\sqrt{\varepsilon}$ do *not* overlap. Whenever $\omega$ crosses (i.e., goes from immediately above to immediately below) one of these four branch-cuts, the refractive index $n(\omega)$ is multiplied by $-1$. Note also that, in the limit when $|\omega| \to \infty$ (along any straight line originating at $\omega = 0$), the complex entities $\mu(\omega)$, $\varepsilon(\omega)$, and the refractive index $n(\omega)$ will all approach 1.0, while $1 - n^2(\omega)$ approaches $(\omega_{pm}^2 + \omega_{pe}^2)/\omega^2$.

Now, with reference to Fig.1, consider a plane-wave of frequency $\omega_q$ and amplitude $E_{1q}$ propagating along the $z$-axis in the free-space region outside the cavity, while the EM field inside the cavity is given by

$$\boldsymbol{E}(z, t) = 2iE_{0q}\hat{\boldsymbol{x}} \sin[n(\omega_q)\omega_q z/c] \exp(-i\omega_q t), \tag{11a}$$

$$\boldsymbol{H}(z, t) = 2Z_0^{-1}[n(\omega_q)/\mu(\omega_q)]E_{0q}\hat{\boldsymbol{y}} \cos[n(\omega_q)\omega_q z/c] \exp(-i\omega_q t). \tag{11b}$$

In the absence of an incident beam, the matching of boundary conditions at $z = d$ yields

$$2iE_{0q} \sin[n(\omega_q)\omega_q d/c] = E_{1q}\exp(i\omega_q d/c), \quad (12a)$$

$$2Z_0^{-1}[n(\omega_q)/\mu(\omega_q)]E_{0q} \cos[n(\omega_q)\omega_q d/c] = Z_0^{-1}E_{1q}\exp(i\omega_q d/c). \quad (12b)$$

The above equations are simultaneously satisfied if and only if $\omega_q$ happens to be a zero of the following function:

$$F(\omega) = n(\omega) \cos[n(\omega)\omega d/c] - i\mu(\omega) \sin[n(\omega)\omega d/c]. \quad (13)$$

We expect the zeros $\omega_q$ of $F(\omega)$ to be confined to the lower-half of the complex $\omega$-plane, because, when the incident beam is set to zero, the time-dependence factor $\exp(-i\omega_q t)$ of the corresponding leaky modes inside the cavity can only decrease with time. Also, considering that $n(-\omega_q^*) = n^*(\omega_q)$ and $\mu(-\omega_q^*) = \mu^*(\omega_q)$, the zeros of $F(\omega)$ always appear in pairs such as $\omega_q$ and $-\omega_q^*$. Trivial leaky modes occur at $\omega_q = \Omega_{1m}$ and $\Omega_{1e}$ (with their twins occurring at $-\omega_q^* = \Omega_{2m}$ and $\Omega_{2e}$), where $n(\Omega_{1,2}) = 0$. Substitution into Eq.(11) reveals that, for these trivial leaky modes, which are associated with the zeros of the refractive index $n(\omega)$, both $E$ and $H$ fields inside and outside the cavity vanish. Finally, with reference to the complex $\omega$-plane of Fig.2, note that when $\omega$ crosses (i.e., moves from immediately above to immediately below) one of the branch-cuts, $n(\omega)$ gets multiplied by $-1$, which causes $F(\omega)$ of Eq.(13) to switch sign.

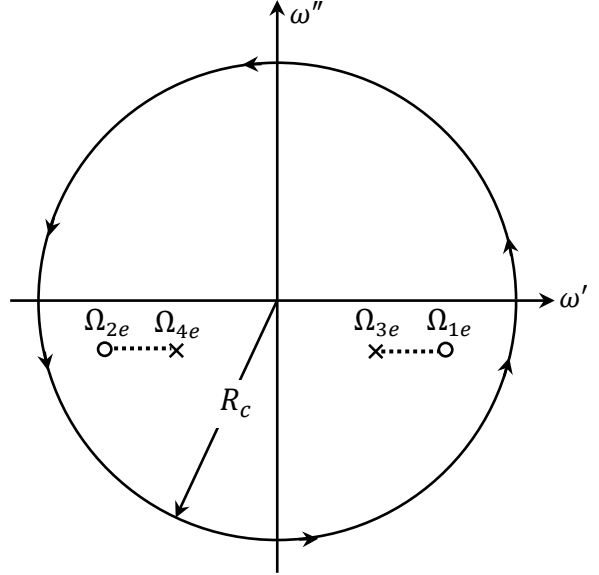

**Fig.2**. Locations in the $\omega$-plane of the poles and zeros of $\varepsilon(\omega)$, whose square root contributes to the refractive index $n(\omega)$ in accordance with Eq.(10). A similar set of poles and zeros, albeit at different locations in the $\omega$-plane, represents $\mu(\omega)$. The dashed lines connecting pairs of adjacent poles and zeros constitute branch-cuts for the function $n(\omega)$. In accordance with the Cauchy-Goursat theorem [6], the integral of a meromorphic function, such as $f(\omega)$, over a circle of radius $R_c$ is $2\pi i$ times the sum of the residues of the function at the poles of $f(\omega)$ that reside within the circle.

Our goal is to express an initial field distribution inside the cavity at $t = t_0$, say,

$$\boldsymbol{E}(z,t_0) = 2iE_0\hat{\boldsymbol{x}} \sin[n(\omega_0)\omega_0 z/c] \exp(-i\omega_0 t_0), \quad (14a)$$

$$\boldsymbol{H}(z,t_0) = 2Z_0^{-1}[n(\omega_0)/\mu(\omega_0)]E_0\hat{\boldsymbol{y}} \cos[n(\omega_0)\omega_0 z/c] \exp(-i\omega_0 t_0); \quad (0 \leq z \leq d), \quad (14b)$$

as a superposition of leaky modes that decay with the passage of time. To this end, we construct the function $G(\omega)$ which incorporates the $E$-field profile, namely, $\sin[n(\omega)\omega z/c]$, the real-valued frequency $\omega_0$ of the initial distribution, and the function $F(\omega)$ of Eq.(13), as follows:

$$G(\omega) = \frac{\sin[n(\omega)\omega z/c]}{(\omega - \omega_0)F(\omega)} = \frac{\sin[n(\omega)\omega z/c]}{(\omega - \omega_0)\{n(\omega) \cos[n(\omega)\omega d/c] - i\mu(\omega) \sin[n(\omega)\omega d/c]\}}. \quad (15)$$

Note that, when $\omega$ crosses a branch-cut, both the numerator and the denominator of $G(\omega)$ switch signs, thus ensuring that the function as a whole remains free of branch-cuts. Let us now examine the behavior of $G(\omega)$ around a large circle of radius $R_c$ centered at the origin of the $\omega$- plane, such as that in Fig.2. Since $\mu(\omega) \to 1 - (\omega_{pm}/\omega)^2$ and $n(\omega) \to 1 - \frac{1}{2}(\omega_{pm}^2 + \omega_{pe}^2)/\omega^2$ everywhere on the circle as $R_c \to \infty$, the limit of $G(\omega)$ will be

$$\lim_{R_c \to \infty} G(\omega) = \lim_{|\omega| \to \infty} \frac{\exp(in\omega z/c) - \exp(-in\omega z/c)}{i(\omega - \omega_0)[(n-\mu)\exp(in\omega d/c) + (n+\mu)\exp(-in\omega d/c)]} = 0. \tag{16}$$

Recognizing that, for all points within the cavity (i.e., $z < d$), the function $G(\omega)$ approaches zero exponentially as $|\omega| \to \infty$, we conclude that the integral of $G(\omega)$ over a large circle of radius $R_c$ vanishes. The Cauchy-Goursat theorem of complex analysis [6] then ensures that all the residues of $G(\omega)$ in the complex-plane add up to zero, that is,

$$\frac{\sin[n(\omega_0)\omega_0 z/c]}{F(\omega_0)} + \sum_q \frac{\sin[n(\omega_q)\omega_q z/c]}{(\omega_q - \omega_0)F'(\omega_q)} = 0. \tag{17}$$

Consequently, the expansion of the initial $E$-field profile inside the cavity expressed as a sum over all the leaky modes is given by

$$\sin[n(\omega_0)\omega_0 z/c] = \sum_q \frac{F(\omega_0)}{(\omega_0 - \omega_q)F'(\omega_q)} \sin[n(\omega_q)\omega_q z/c]. \tag{18}$$

A similar method may be used to arrive at an expansion of the initial $H$-field distribution in terms of the same leaky modes as in Eq.(18). In this case, the function $G(\omega)$ must be chosen as

$$G(\omega) = \frac{n(\omega)\cos[n(\omega)\omega z/c]}{(\omega - \omega_0)\mu(\omega)F(\omega)}. \tag{19}$$

Once again, since the integral of the above $G(\omega)$ around a large circle of radius $R_c$ (centered at the origin) approaches zero, the residues of $G(\omega)$ in the present case also must add up to zero. Consequently, the procedure for expanding the initial $H$-field distribution via Eq.(19) is similar to that used previously to expand the initial $E$-field via Eq.(15).

**4. Numerical results**. Figure 3 shows the ratio of the $E$-field amplitude inside a dielectric slab to the incident $E$-field, plotted versus the excitation frequency $\omega$ normalized by $\omega_0 = 1.885 \times 10^{15}$ rad/sec (corresponding to the free space wavelength $\lambda_0 = 1.0$ μm). The 500 nm-thick slab has refractive index $n = 3.75 + 0.0116i$ at $\omega = \omega_0$, and, as shown in Fig.1, is coated on one of its facets by a perfect reflector. It is assumed that $\mu(\omega) = 1.0$, and that the permittivity $\varepsilon(\omega)$ follows a single Lorentz oscillator model with resonance frequency $\omega_r = 4\omega_0$, damping coefficient $\gamma = 0.1\omega_0$, and plasma frequency $\omega_p = 14\omega_0$. In the interval between the pole and zero of the refractive index, namely, $[\Omega_{3e}, \Omega_{1e}]$ (see Fig.2), the field amplitude inside the cavity is seen to be vanishingly small. Outside this "forbidden" range of frequencies, the field has resonance peaks at specific frequencies, and the $E_{\text{inside}}/E_{\text{incident}}$ ratio between adjacent peaks and valleys can vary by as much as a factor of 4.

The contour plots of Fig.4 show, within the complex $\omega$-plane, the zeros of $\text{Re}[F(\omega)]$ in red and the zeros of $\text{Im}[F(\omega)]$ in blue. Both the real and imaginary parts of $\omega$ are normalized by the reference frequency $\omega_0$. The points where the contours cross each other—several of them circled in the plot—represent the zeros of $F(\omega)$, which we have denoted by $\omega_q = \omega_q' + i\omega_q''$ and referred to as leaky-mode frequencies. The resonance peaks seen in Fig.3 occur at or near the frequencies $\omega = \omega_q'$ of the various leaky modes. The region of the $\omega$-plane depicted in Fig.4(a) contains all the leaky-mode frequencies to the left of $\Omega_{3e}$; a large number of such frequencies are seen to accumulate in the vicinity of $\omega = \Omega_{3e}$, where the coupling of the incident light to the cavity is weak, and the damping within the slab is dominated by absorption losses. The region of the $\omega$-plane depicted in Fig.4(b) contains all the leaky-mode frequencies to the right of $\Omega_{1e}$. The imaginary part $\omega_q''$ of these frequencies acquires large negative values as the real part $\omega_q'$ of the corresponding leaky frequency increases. No leaky modes reside in the upper-half of the $\omega$-plane, nor are there any leaky modes in the strip between $\Omega_{1e}$ and $\Omega_{3e}$.

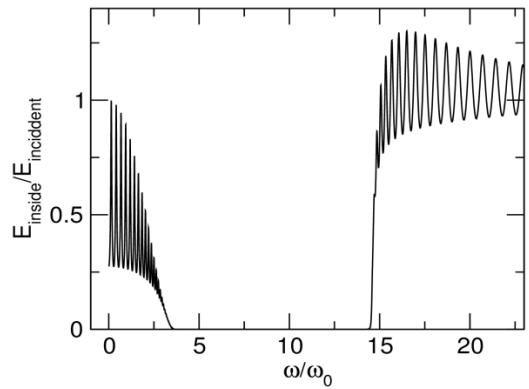

**Fig. 3**. Ratio of the $E$-field amplitude inside a dielectric slab to the incident $E$-field, plotted versus normalized frequency $\omega/\omega_0$.

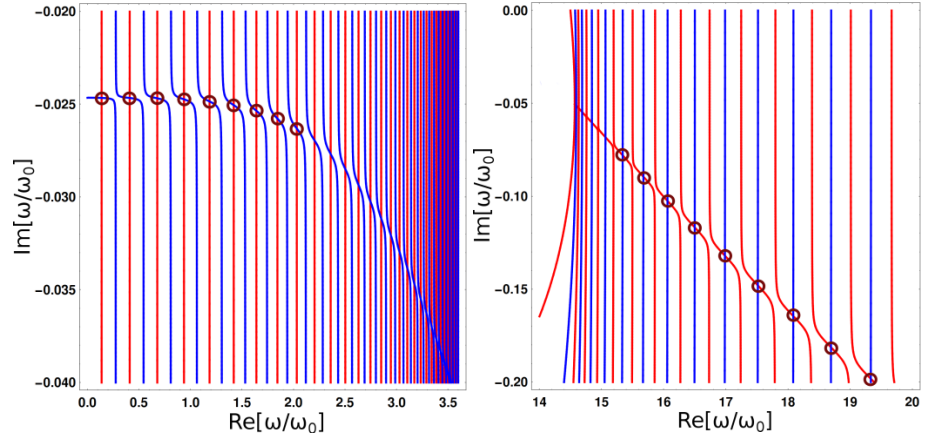

**Fig. 4**. Contour plots in the $\omega$-plane, showing the zeros of $\text{Re}[F(\omega)]$ in red and the zeros of $\text{Im}[F(\omega)]$ in blue. Both the real and imaginary parts of $\omega$ are normalized by the reference frequency $\omega_0$. (a) Region of $\omega$-plane to the left of $\Omega_{3e}$. (b) Region of $\omega$-plane to the right of $\Omega_{1e}$. The points where the contours cross each other (several of them circled) are the zeros of $F(\omega)$.

A comparison of Figs.3 and 4 reveals the close relationship between the leaky mode frequencies and the resonances of the dielectric slab. Resonances occur at or near the frequencies $\omega = \omega_q'$, and the height and width of a resonance line are, by and large, determined by the decay rate $\omega_q''$ of the corresponding leaky mode—unless the leaky mode frequency happens to be so close to the pole(s) of the refractive index $n(\omega)$ that the strong absorption within the medium would suppress the resonance. It must be emphasized that the presence of a gap in the frequency domain (such as that between $\omega = \Omega_{3e}$ and $\omega = \Omega_{1e}$ in the present example) should not prevent the leaky modes from forming a basis, because, as an ensemble, the leaky modes are expected (on physical grounds) to carry all the spatial frequencies needed to capture the various features of arbitrary initial $E$-field and $H$-field distributions.

**5. Leaky modes propagating at oblique angle relative to the surface normal**. The diagram in Fig.5 shows a leaky mode of a dielectric slab, whose $k$-vector has a component $k_x$ along the $x$-axis. Here $k_x$ is assumed to be a real-valued and positive constant. Although the following discussion is confined to the case of transverse magnetic (TM) polarization, the analysis is straightforward and can be readily extended to the case of transverse electric (TE) polarization as well. Inside the slab depicted in Fig.5, $k_{z0} = \sqrt{\mu\varepsilon(\omega/c)^2 - k_x^2}$ while the $H$ and $E$ field amplitudes are $H_{0q}\hat{y}$ and $E_{0q} = -(k_x\hat{x} \pm k_{z0}\hat{z}) \times H_{0q}\hat{y}/(\varepsilon_0\varepsilon\omega)$, respectively. The total EM field inside the slab is given by

$$\boldsymbol{H}_{\text{in}}(\boldsymbol{r},t) = H_0\hat{y}\{\exp[i(k_x x - k_{z0}z)] + \exp[i(k_x x + k_{z0}z)]\}\exp(-i\omega t). \tag{20a}$$

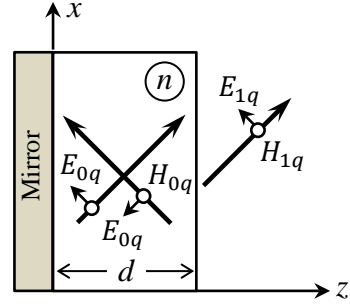

**Fig. 5**. Dielectric slab of thickness $d$ and refractive index $n(\omega) = \sqrt{\mu\varepsilon}$, supporting a TM polarized leaky mode. Inside the slab, the $k$-vectors of the counter-propagating plane-waves are $k_x\hat{x} \pm k_{z0}\hat{z}$, while their field amplitudes are $\boldsymbol{H} = H_{0q}\hat{y}$ and $\boldsymbol{E} = (\pm k_{z0}\hat{x} - k_x\hat{z})H_{0q}/(\varepsilon_0\varepsilon\omega)$. The plane-wave that leaves the slab resides in free space; its $k$-vector is $k_x\hat{x} + k_{z1}\hat{z}$, and its field amplitudes are $\boldsymbol{H} = H_{1q}\hat{y}$ and $\boldsymbol{E} = (k_{z1}\hat{x} - k_x\hat{z})H_{1q}/(\varepsilon_0\omega)$.

$$\boldsymbol{E}_{\text{in}}(\boldsymbol{r},t) = [H_0/\varepsilon_0\varepsilon(\omega)\omega]\{-(k_{z0}\hat{x} + k_x\hat{z})\exp[\mathrm{i}(k_x x - k_{z0}z)]$$
$$+ (k_{z0}\hat{x} - k_x\hat{z})\exp[\mathrm{i}(k_x x + k_{z0}z)]\}\exp(-\mathrm{i}\omega t). \qquad (20\text{b})$$

Outside the slab, $k_{z1} = \sqrt{(\omega/c)^2 - k_x^2}$, $\boldsymbol{H}_{1q} = H_{1q}\hat{y}$ and $\boldsymbol{E}_{1q} = -(k_x\hat{x} + k_{z1}\hat{z}) \times H_{1q}\hat{y}/(\varepsilon_0\omega)$. The total EM field in the free-space region outside the slab, where $z > d$, is thus given by

$$\boldsymbol{H}_{\text{out}}(\boldsymbol{r},t) = H_1\hat{y}\exp[\mathrm{i}(k_x x + k_{z1}z)]\exp(-\mathrm{i}\omega t). \qquad (21\text{a})$$

$$\boldsymbol{E}_{\text{out}}(\boldsymbol{r},t) = (H_1/\varepsilon_0\omega)(k_{z1}\hat{x} - k_x\hat{z})\exp[\mathrm{i}(k_x x + k_{z1}z)]\exp(-\mathrm{i}\omega t). \qquad (21\text{b})$$

Matching the boundary conditions at $z = d$, we find

$$H_0[\exp(-\mathrm{i}k_{z0}d) + \exp(\mathrm{i}k_{z0}d)] = H_1\exp(\mathrm{i}k_{z1}d). \qquad (22\text{a})$$

$$k_{z0}H_0[-\exp(-\mathrm{i}k_{z0}d) + \exp(\mathrm{i}k_{z0}d)] = \varepsilon(\omega)k_{z1}H_1\exp(\mathrm{i}k_{z1}d). \qquad (22\text{b})$$

The above equations are simultaneously satisfied when the frequency $\omega$ satisfies the following characteristic equation:

$$F(\omega) = \varepsilon(\omega)\sqrt{(\omega/c)^2 - k_x^2}\cos[\sqrt{\mu\varepsilon(\omega/c)^2 - k_x^2}\,d]$$
$$- \mathrm{i}\sqrt{\mu\varepsilon(\omega/c)^2 - k_x^2}\sin[\sqrt{\mu\varepsilon(\omega/c)^2 - k_x^2}\,d] = 0. \qquad (23)$$

As before, the zeros of $F(\omega)$ represent the leaky-mode frequencies associated with initial conditions whose dependence on the $x$-coordinate is given by the phase-factor $\exp(\mathrm{i}k_x x)$. In the $\omega$-plane, the function $F(\omega)$ of Eq.(23) has a branch-cut on the real-axis between the zeros of $k_{z1}$, i.e., $\pm ck_x$. It also has other branch-cuts associated with the poles and zeros of $k_{z0}$, as was the case for normal incidence discussed in Sec.3. The branch-cut on the real-axis is troublesome, as it cannot be easily eliminated in order to render $F(\omega)$ analytic. While, on physical grounds, we believe that an arbitrary initial distribution can still be expressed as a superposition of leaky modes whose frequencies are the roots of Eq.(23), the aforementioned branch-cut residing on the real-axis in the $\omega$-plane prevents us from proving the completeness of such leaky modes. We must remain content with the fact that the computed roots of Eq.(23) reside in the lower-half of the $\omega$-plane, and that these roots have properties that are expected of leaky modes whose $x$-dependence is given by $\exp(\mathrm{i}k_x x)$. The quest for a completeness proof, however, must continue, and the methods of the preceding sections, which worked so well in the case of normal incidence on a dielectric slab, must somehow be extended to encompass the case of oblique incidence.

**6. Leaky modes of a solid dielectric sphere**. The vector spherical harmonics of the EM field within a homogeneous, isotropic, linear medium having permeability $\mu_0\mu(\omega)$ and permittivity $\varepsilon_0\varepsilon(\omega)$ are found by solving Maxwell's equations in spherical coordinates [7]. The electric and magnetic field profiles for Transverse Electric (TE) and Transverse Magnetic (TM) modes of the EM field are found to be[†]

$m = 0$ TE mode ($E_r = 0$):

$$\boldsymbol{E}(\boldsymbol{r},t) = \frac{E_0 J_{\ell+\frac{1}{2}}(kr)}{\sqrt{kr}} P_\ell^1(\cos\theta)\exp(-i\omega t)\widehat{\boldsymbol{\varphi}}. \tag{24}$$

$$\boldsymbol{H}(\boldsymbol{r},t) = \frac{E_0}{\mu_0\mu(\omega)r\omega}\left\{\frac{J_{\ell+\frac{1}{2}}(kr)}{i\sqrt{kr}}[\cot\theta\, P_\ell^1(\cos\theta) - \sin\theta\,\dot{P}_\ell^1(\cos\theta)]\widehat{\boldsymbol{r}}\right.$$
$$\left. - \frac{kr\dot{J}_{\ell+\frac{1}{2}}(kr)+\frac{1}{2}J_{\ell+\frac{1}{2}}(kr)}{i\sqrt{kr}} P_\ell^1(\cos\theta)\widehat{\boldsymbol{\theta}}\right\}\exp(-i\omega t). \tag{25}$$

$m \neq 0$ TE mode ($E_r = 0$):

$$\boldsymbol{E}(\boldsymbol{r},t) = E_0\left[\frac{J_{\ell+\frac{1}{2}}(kr)}{\sqrt{kr}}\frac{P_\ell^m(\cos\theta)}{\sin\theta}\widehat{\boldsymbol{\theta}} + \frac{J_{\ell+\frac{1}{2}}(kr)}{im\sqrt{kr}}\sin\theta\,\dot{P}_\ell^m(\cos\theta)\widehat{\boldsymbol{\varphi}}\right]\exp[i(m\varphi - \omega t)]. \tag{26}$$

$$\boldsymbol{H}(\boldsymbol{r},t) = -\frac{E_0}{\mu_0\mu(\omega)r\omega}\left\{\frac{\ell(\ell+1)J_{\ell+\frac{1}{2}}(kr)}{m\sqrt{kr}}P_\ell^m(\cos\theta)\widehat{\boldsymbol{r}} - \frac{kr\dot{J}_{\ell+\frac{1}{2}}(kr)+\frac{1}{2}J_{\ell+\frac{1}{2}}(kr)}{m\sqrt{kr}}\sin\theta\,\dot{P}_\ell^m(\cos\theta)\widehat{\boldsymbol{\theta}}\right.$$
$$\left. - \frac{kr\dot{J}_{\ell+\frac{1}{2}}(kr)+\frac{1}{2}J_{\ell+\frac{1}{2}}(kr)}{i\sqrt{kr}}\frac{P_\ell^m(\cos\theta)}{\sin\theta}\widehat{\boldsymbol{\varphi}}\right\}\exp[i(m\varphi - \omega t)]. \tag{27}$$

$m = 0$ TM mode ($H_r = 0$):

$$\boldsymbol{E}(\boldsymbol{r},t) = -\frac{H_0}{\varepsilon_0\varepsilon(\omega)r\omega}\left\{\frac{J_{\ell+\frac{1}{2}}(kr)}{i\sqrt{kr}}[\cot\theta\, P_\ell^1(\cos\theta) - \sin\theta\,\dot{P}_\ell^1(\cos\theta)]\widehat{\boldsymbol{r}}\right.$$
$$\left. - \frac{kr\dot{J}_{\ell+\frac{1}{2}}(kr)+\frac{1}{2}J_{\ell+\frac{1}{2}}(kr)}{i\sqrt{kr}} P_\ell^1(\cos\theta)\widehat{\boldsymbol{\theta}}\right\}\exp(-i\omega t). \tag{28}$$

$$\boldsymbol{H}(\boldsymbol{r},t) = \frac{H_0 J_{\ell+\frac{1}{2}}(kr)}{\sqrt{kr}} P_\ell^1(\cos\theta)\exp(-i\omega t)\widehat{\boldsymbol{\varphi}}. \tag{29}$$

$m \neq 0$ TM mode ($H_r = 0$):

$$\boldsymbol{E}(\boldsymbol{r},t) = \frac{H_0}{\varepsilon_0\varepsilon(\omega)r\omega}\left\{\frac{\ell(\ell+1)J_{\ell+\frac{1}{2}}(kr)}{m\sqrt{kr}}P_\ell^m(\cos\theta)\widehat{\boldsymbol{r}} - \frac{kr\dot{J}_{\ell+\frac{1}{2}}(kr)+\frac{1}{2}J_{\ell+\frac{1}{2}}(kr)}{m\sqrt{kr}}\sin\theta\,\dot{P}_\ell^m(\cos\theta)\widehat{\boldsymbol{\theta}}\right.$$
$$\left. - \frac{kr\dot{J}_{\ell+\frac{1}{2}}(kr)+\frac{1}{2}J_{\ell+\frac{1}{2}}(kr)}{i\sqrt{kr}}\frac{P_\ell^m(\cos\theta)}{\sin\theta}\widehat{\boldsymbol{\varphi}}\right\}\exp[i(m\varphi - \omega t)]. \tag{30}$$

$$\boldsymbol{H}(\boldsymbol{r},t) = H_0\left[\frac{J_{\ell+\frac{1}{2}}(kr)}{\sqrt{kr}}\frac{P_\ell^m(\cos\theta)}{\sin\theta}\widehat{\boldsymbol{\theta}} + \frac{J_{\ell+\frac{1}{2}}(kr)}{im\sqrt{kr}}\sin\theta\,\dot{P}_\ell^m(\cos\theta)\widehat{\boldsymbol{\varphi}}\right]\exp[i(m\varphi - \omega t)]. \tag{31}$$

In the above equations, the Bessel function $J_\nu(z)$ and its derivative with respect to $z$, $\dot{J}_\nu(z)$, could be replaced by a Bessel function of the second kind, $Y_\nu(z)$, and its derivative, $\dot{Y}_\nu(z)$, or by Hankel functions of type 1 or type 2, namely, $\mathcal{H}_\nu^{(1,2)}(z)$, and corresponding derivatives $\dot{\mathcal{H}}_\nu^{(1,2)}(z)$.

---

[†] For a given $m$, the TM mode may be obtained from the corresponding TE mode by substituting $\boldsymbol{E}$ for $\boldsymbol{H}$, and $-\boldsymbol{H}$ for $\boldsymbol{E}$, keeping in mind that $r\omega = kr/\sqrt{\mu_0\varepsilon_0\mu(\omega)\varepsilon(\omega)}$, and that the $E/H$ amplitude ratio for each mode is always given by $\sqrt{\mu_0\mu(\omega)/\varepsilon_0\varepsilon(\omega)}$.

The (complex) field amplitudes are denoted by $E_0$ and $H_0$. In our spherical coordinate system, the point $\boldsymbol{r}$ is at a distance $r$ from the origin, its polar and azimuthal angles being $\theta$ and $\varphi$. The oscillation frequency is $\omega$, and the wave-number $k$ is defined as $k(\omega) = n(\omega)k_0$, where $k_0 = \omega/c$, and $n(\omega) = \sqrt{\mu(\omega)\varepsilon(\omega)}$ is the refractive index of the host medium. The integers $\ell \geq 1$, and $m$ (ranging from $-\ell$ to $+\ell$) specify the polar and azimuthal mode numbers. $P_\ell^m(\zeta)$ is an associated Legendre function, while $\dot{P}_\ell^m(\zeta)$ is its derivative with respect to $\zeta$. Finally, the various Bessel functions of half-integer order are defined by the following formulas [8]:

$$J_{\ell+\frac{1}{2}}(z) = \sqrt{\frac{2}{\pi z}} \left\{ \sin(z - \tfrac{1}{2}\ell\pi) \sum_{k=0}^{\lfloor \ell/2 \rfloor} \frac{(-1)^k(\ell+2k)!}{(2k)!(\ell-2k)!} \left(\frac{1}{2z}\right)^{2k} \right. \\ \left. + \cos(z - \tfrac{1}{2}\ell\pi) \sum_{k=0}^{\lfloor (\ell-1)/2 \rfloor} \frac{(-1)^k(\ell+2k+1)!}{(2k+1)!(\ell-2k-1)!} \left(\frac{1}{2z}\right)^{2k+1} \right\}. \quad \leftarrow \boxed{\text{G\&R 8.461-1}} \quad (32)$$

$$Y_{\ell+\frac{1}{2}}(z) = (-1)^{\ell-1}\sqrt{\frac{2}{\pi z}} \left\{ \cos(z + \tfrac{1}{2}\ell\pi) \sum_{k=0}^{\lfloor \ell/2 \rfloor} \frac{(-1)^k(\ell+2k)!}{(2k)!(\ell-2k)!} \left(\frac{1}{2z}\right)^{2k} \right. \\ \left. - \sin(z + \tfrac{1}{2}\ell\pi) \sum_{k=0}^{\lfloor (\ell-1)/2 \rfloor} \frac{(-1)^k(\ell+2k+1)!}{(2k+1)!(\ell-2k-1)!} \left(\frac{1}{2z}\right)^{2k+1} \right\}. \quad \leftarrow \boxed{\begin{array}{l}\text{G\&R 8.461-2}\\\text{G\&R 8.465-1}\end{array}} \quad (33)$$

$$\mathcal{H}^{(1)}_{\ell+\frac{1}{2}}(z) = \sqrt{\frac{2}{\pi z}} \exp\{i[z - \tfrac{1}{2}(\ell+1)\pi]\} \sum_{k=0}^{\ell} \frac{(\ell+k)!}{k!(\ell-k)!} \left(\frac{i}{2z}\right)^k. \quad \leftarrow \boxed{\text{G\&R 8.466-1}} \quad (34)$$

Note that $\sqrt{z}J_{\ell+\frac{1}{2}}(z)$ is an even function of $z$ when $\ell = 1, 3, 5, \cdots$, and an odd function when $\ell = 2, 4, 6, \cdots$. This fact will be needed later on, when we try to argue that certain branch-cuts in the complex $\omega$-plane are inconsequential.

Consider now a solid dielectric sphere of radius $R$, relative permeability $\mu(\omega)$, and relative permittivity $\varepsilon(\omega)$. Inside the particle, the radial dependence of the TE mode is governed by a Bessel function of the first kind, $E_0 J_{\ell+\frac{1}{2}}(kr)$, and its derivative. The refractive index of the spherical particle being $n(\omega) = \sqrt{\mu(\omega)\varepsilon(\omega)}$, the corresponding wave-number inside the particle is $k(\omega) = n(\omega)k_0 = n(\omega)\omega/c$. The particle is surrounded by free space, which is host to an *outgoing* spherical harmonic whose radial dependence is governed by a type 1 Hankel function, $E_1 \mathcal{H}^{(1)}_{\ell+\frac{1}{2}}(k_0 r)$, and its derivative. Invoking the Bessel function identity $z\dot{J}_\nu(z) = \nu J_\nu(z) - zJ_{\nu+1}(z)$ — which applies to $Y_\nu(z)$ and $\mathcal{H}^{(1,2)}_\nu(z)$ as well — we find, upon matching the boundary conditions at $r = R$, that the following two equations must be simultaneously satisfied:

$$\frac{E_0 J_{\ell+\frac{1}{2}}(nk_0 R)}{\sqrt{nk_0 R}} = \frac{E_1 \mathcal{H}^{(1)}_{\ell+\frac{1}{2}}(k_0 R)}{\sqrt{k_0 R}}, \tag{35a}$$

$$\frac{E_0[(\ell+1)J_{\ell+\frac{1}{2}}(nk_0 R) - nk_0 R J_{\ell+3/2}(nk_0 R)]}{\mu(\omega)\sqrt{nk_0 R}} = \frac{E_1[(\ell+1)\mathcal{H}^{(1)}_{\ell+\frac{1}{2}}(k_0 R) - k_0 R \mathcal{H}^{(1)}_{\ell+3/2}(k_0 R)]}{\sqrt{k_0 R}}. \tag{35b}$$

Streamlining the above equations, we arrive at

$$\begin{bmatrix} J_{\ell+\frac{1}{2}}(nk_0R) & -\sqrt{n}\mathcal{H}^{(1)}_{\ell+\frac{1}{2}}(k_0R) \\ (\ell+1)J_{\ell+\frac{1}{2}}(nk_0R) - nk_0RJ_{\ell+3/2}(nk_0R) & -\mu\sqrt{n}[(\ell+1)\mathcal{H}^{(1)}_{\ell+\frac{1}{2}}(k_0R) - k_0R\mathcal{H}^{(1)}_{\ell+3/2}(k_0R)] \end{bmatrix} \begin{bmatrix} E_0 \\ E_1 \end{bmatrix} = 0. \quad (36)$$

A non-trivial solution for $E_0$ and $E_1$ thus exists if and only if the determinant of the coefficient matrix in Eq.(36) vanishes, that is,

$$F(\omega) = nk_0R\mathcal{H}^{(1)}_{\ell+\frac{1}{2}}(k_0R)J_{\ell+3/2}(nk_0R) + [(\mu-1)(\ell+1)\mathcal{H}^{(1)}_{\ell+\frac{1}{2}}(k_0R) - \mu k_0R\mathcal{H}^{(1)}_{\ell+3/2}(k_0R)]J_{\ell+\frac{1}{2}}(nk_0R) = 0. \quad (37)$$

This is the characteristic equation for leaky TE modes, whose solutions comprise the entire set of leaky frequencies $\omega_q$. The index $q$ is used here to enumerate the various leaky-mode frequencies. For TM modes, $\mu(\omega)$ in Eq.(37) must be replaced by $\varepsilon(\omega)$. Equation (37) must be solved numerically for complex frequencies $\omega_q$; these being characteristic frequencies of the particle's leaky modes, one expects (on physical grounds) to find all the roots $\omega_q$ of $F(\omega)$ in the lower-half of the complex plane. Note that $\sqrt{n}F(\omega)$ is an even function of $n$ when $\ell = 1, 3, 5, \cdots$, and an odd function when $\ell = 2, 4, 6, \cdots$. This is because successive Bessel functions $J_{\ell+\frac{1}{2}}$ and $J_{\ell+3/2}$ alternate between odd and even parities. Note also that $F(\omega)$ vanishes at the zeros of $n(\omega)$, that is, $F(\Omega_1) = F(\Omega_2) = 0$. Finally, when $\omega \to 0$, $F(\omega)$ approaches a constant (see the Appendix), and when $|\omega| \to \infty$, $\mu(\omega) \to 1 - (\omega_{pm}/\omega)^2$ and $\varepsilon(\omega) \to 1 - (\omega_{pe}/\omega)^2$, thus allowing the asymptotic behavior of $F(\omega)$ to be determined from Eqs.(32) and (34).

Our goal is to express an initial field distribution inside the particle (e.g., one of the spherical harmonic waveforms given by Eqs.(24)-(31), which oscillate at a real-valued frequency $\omega_0$) as a superposition of leaky modes, each having its own complex frequency $\omega_q$. To this end, we must form a meromorphic function $G(\omega)$ incorporating the following attributes:

i) The function $F(\omega)$ of Eq.(37) appears in the denominator of $G(\omega)$, thus causing the zeros of $F(\omega)$ to act as poles for $G(\omega)$.

ii) A desired initial waveform, say, $J_{\ell+\frac{1}{2}}[\omega n(\omega)r/c]$, appearing in the numerator of $G(\omega)$.

iii) The real-valued frequency $\omega_0$ associated with the initial waveform acting as a pole for $G(\omega)$.

iv) In the limit when $|\omega| \to \infty$, $G(\omega) \to 0$ exponentially, so that $\oint G(\omega)d\omega$ over a circle of large radius $R_c$ vanishes.

A simple (although by no means the only) such function is

$$G(\omega) = \frac{\omega^{3/2}\exp(iR\omega/c)J_{\ell+\frac{1}{2}}(kr)}{(\omega-\omega_0)F(\omega)}. \quad (38)$$

With reference to Eq.(32), note that the pre-factor $1/\sqrt{n}$ of the Bessel function in the numerator of $G(\omega)$ cancels the corresponding pre-factor that accompanies the denominator. The remaining part of the Bessel function in the numerator will then have the same parity with respect to $n(\omega)$ as the function that appears in the denominator. Consequently, switching the sign of $n(\omega)$ does *not* alter $G(\omega)$, indicating that the branch-cuts associated with $n(\omega)$ in the complex $\omega$-plane do *not* introduce discontinuities into $G(\omega)$. The presence of $\sqrt{\omega}\exp(iR\omega/c)$ in the numerator of $G(\omega)$ is intended to eliminate the undesirable features of the Hankel functions appearing in the denominator. The function $G(\omega)$ is thus analytic everywhere except at the poles, where its denominator vanishes. The poles, of course, consist of $\omega = \omega_0$, which is the frequency of the initial EM field residing inside the spherical particle at $t = 0$, and $\omega = \omega_q$, which are the leaky-mode frequencies found by solving Eq.(37)—or its TM mode counterpart.

In the limit $|\omega| \to \infty$, where $\mu(\omega) \to 1 - (\omega_{pm}/\omega)^2$ and $\varepsilon(\omega) \to 1 - (\omega_{pe}/\omega)^2$, we find that $G(\omega)$ approaches zero exponentially. Thus, the vanishing of $\oint G(\omega)d\omega$ around a circle of large radius $R_c$ means that all the residues of $G(\omega)$ must add up to zero, that is,

$$\frac{\omega_0^{3/2}\exp(iR\omega_0/c)J_{\ell+\frac{1}{2}}[\omega_0 n(\omega_0)r/c]}{F(\omega_0)} + \sum_q \frac{\omega_q^{3/2}\exp(iR\omega_q/c)J_{\ell+\frac{1}{2}}[\omega_q n(\omega_q)r/c]}{(\omega_q - \omega_0)F'(\omega_q)} = 0. \tag{39}$$

The initial field distribution $J_{\ell+\frac{1}{2}}[\omega_0 n(\omega_0)r/c]$ may thus be expanded as the following superposition of all the leaky modes:

$$J_{\ell+\frac{1}{2}}[\omega_0 n(\omega_0)r/c] = \sum_q \frac{\omega_q^{3/2}\exp[iR(\omega_q - \omega_0)/c]F(\omega_0)}{\omega_0^{3/2}(\omega_0 - \omega_q)F'(\omega_q)} \times J_{\ell+\frac{1}{2}}[\omega_q n(\omega_q)r/c]. \tag{40}$$

To incorporate into the initial distribution the denominator $\sqrt{kr}$, which accompanies all the field components in Eqs.(24)-(31), we modify Eq.(40)—albeit trivially—as follows:

$$\frac{J_{\ell+\frac{1}{2}}[\omega_0 n(\omega_0)r/c]}{\sqrt{\omega_0 n(\omega_0)r/c}} = \sum_q \frac{\omega_q^{3/2}\sqrt{\omega_q n(\omega_q)}\exp[iR(\omega_q - \omega_0)/c]F(\omega_0)}{\omega_0^{3/2}\sqrt{\omega_0 n(\omega_0)}(\omega_0 - \omega_q)F'(\omega_q)} \times \frac{J_{\ell+\frac{1}{2}}[\omega_q n(\omega_q)r/c]}{\sqrt{\omega_q n(\omega_q)r/c}}. \tag{41}$$

Taking advantage of the flexibility of $G(\omega)$, we now extend the same treatment to the remaining components of the EM field. For instance, if we choose

$$G(\omega) = \frac{\sqrt{\omega}\exp(iR\omega/c)J_{\ell+\frac{1}{2}}(kr)}{(\omega - \omega_0)\mu(\omega)F(\omega)}, \tag{42}$$

then $G(\omega) \to 0$ exponentially in the limit when $|\omega| \to \infty$, resulting in a vanishing integral around the circle of large radius $R_c$ in the $\omega$-plane. We thus arrive at an alternative form of Eq.(41), which is useful for expanding the field component $H_r$ appearing in Eqs.(25) and (27), that is,

$$\frac{J_{\ell+\frac{1}{2}}[\omega_0 n(\omega_0)r/c]}{\mu(\omega_0)r\omega_0\sqrt{\omega_0 n(\omega_0)r/c}} = \sum_q \frac{\omega_q^{3/2}\sqrt{\omega_q n(\omega_q)}\exp[iR(\omega_q - \omega_0)/c]F(\omega_0)}{\omega_0^{3/2}\sqrt{\omega_0 n(\omega_0)}(\omega_0 - \omega_q)F'(\omega_q)} \times \frac{J_{\ell+\frac{1}{2}}[\omega_q n(\omega_q)r/c]}{\mu(\omega_q)r\omega_q\sqrt{\omega_q n(\omega_q)r/c}}. \tag{43}$$

Finally, if we choose

$$G(\omega) = \frac{\sqrt{\omega}\exp(iR\omega/c)[krj'_{\ell+\frac{1}{2}}(kr) + \frac{1}{2}J_{\ell+\frac{1}{2}}(kr)]}{(\omega - \omega_0)\mu(\omega)F(\omega)}, \tag{44}$$

it continues to be meromorphic (i.e., free of branch-cuts), and will have a vanishing integral over a large circle of radius $R_c$ in the limit when $R_c \to \infty$. The relevant expansion of the field components $H_\theta$ and $H_\varphi$ appearing in Eqs.(25) and (27) will then be

$$\frac{[\omega_0 n(\omega_0)r/c]j'_{\ell+\frac{1}{2}}[\omega_0 n(\omega_0)r/c] + \frac{1}{2}J_{\ell+\frac{1}{2}}[\omega_0 n(\omega_0)r/c]}{\mu(\omega_0)r\omega_0\sqrt{\omega_0 n(\omega_0)r/c}}$$

$$= \sum_q \frac{\omega_q^{3/2}\sqrt{\omega_q n(\omega_q)}\exp[iR(\omega_q - \omega_0)/c]F(\omega_0)}{\omega_0^{3/2}\sqrt{\omega_0 n(\omega_0)}(\omega_0 - \omega_q)F'(\omega_q)} \times \frac{[\omega_q n(\omega_q)r/c]j'_{\ell+\frac{1}{2}}[\omega_q n(\omega_q)r/c] + \frac{1}{2}J_{\ell+\frac{1}{2}}[\omega_q n(\omega_q)r/c]}{\mu(\omega_q)r\omega_q\sqrt{\omega_q n(\omega_q)r/c}}. \tag{45}$$

In this way, one can expand into a superposition of leaky modes the various $\boldsymbol{E}$ and $\boldsymbol{H}$ field components that comprise an initial distribution. It will then be possible to follow each leaky mode as its phase evolves while its amplitude decays with the passage of time.

**7. Numerical results for a solid glass sphere**. Figure 6 shows the resonances of a dielectric sphere of radius $R = 50\lambda_0$ and refractive index $n = 1.5$ at and around the reference frequency $\omega_0 = 1.216 \times 10^{15}$ rad/sec, which corresponds to the free-space wavelength $\lambda_0 = 1.55\ \mu m$. In this and subsequent figures, the frequency $\omega$ is normalized by $\omega_0$. The contours of real and imaginary parts of the characteristic equation $F(\omega) = 0$ can be plotted in the complex $\omega$-plane, as was done for a dielectric slab in Fig.4. Where the contours cross each other, the function $F(\omega)$ vanishes, indicating the existence of a leaky mode at the crossing frequency $\omega_q = \omega'_q + i\omega''_q$. The ratio $|\omega'_q|/|\omega''_q|$ is a measure of the $Q$-factor of the spherical cavity at the excitation frequency $\omega = \omega'_q$. Shown in Fig.6 are the computed $Q$-factors of the spherical cavity for both TE and TM modes at the various resonance frequencies corresponding to $\ell = 340$. (Note that the characteristic equation does not depend on $m$, which indicates that, for a given integer $\ell$, the modes associated with all $m$ between $-\ell$ and $\ell$ are degenerate.) The lowest resonance frequency occurs at $\omega \cong 0.78\omega_0$. The large values of $Q$ seen in Fig.6 are a consequence of the fact that the refractive index $n$ is assumed to be purely real; later, when absorption is incorporated into the model via the imaginary part of $n$, the $Q$-factors will drop to more reasonable values.

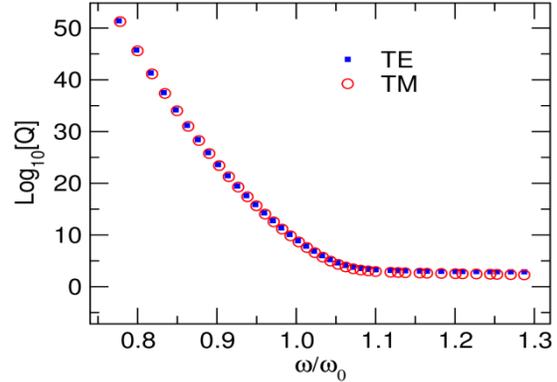

**Fig. 6**. Computed $Q$-factor versus the resonance frequency for a dielectric sphere of radius $R = 50\lambda_0$ and refractive index $n = 1.5$ in the vicinity of $\omega_0 = 1.216 \times 10^{15}$ rad/sec. On the horizontal axis, the frequency $\omega$ is normalized by $\omega_0$. The leaky mode frequencies $\omega_q = \omega'_q + i\omega''_q$ are solutions of $F(\omega) = 0$, which have been found numerically. The ratio $|\omega'_q|/|\omega''_q|$ is used as a measure of the $Q$-factor of the spherical cavity at the excitation frequency $\omega = \omega'_q$. The figure shows computed $Q$-factors for both TE and TM modes at the various resonance frequencies of the dielectric sphere corresponding to $\ell = 340$.

The direct method of determining the resonances of the spherical cavity involves the computation of the ratio $E_{\text{inside}}/E_{\text{incident}}$ for an incident Hankel function of type 2 (incoming wave) and a fixed mode number $\ell$. Once again, the results are independent of the azimuthal mode number $m$, as the modes associated with $m = -\ell$ to $\ell$ are all degenerate. Figure 7 shows plots of $E_{\text{inside}}/E_{\text{incident}}$ for the spherical cavity of radius $R = 50\lambda_0$, refractive index $n = 1.5$, and mode number $\ell = 340$, at and around $\omega_0 = 1.216 \times 10^{15}$ rad/sec; the results for both TE and TM modes are presented in the figure. The resonances are seen to be strong with narrow linewidths. Note that, outside the resonance peaks and especially at lower frequencies, the coupling of the incident beam to the cavity is extremely weak. The TE and TM modes are quite similar in their coupling efficiencies and resonant line-shapes, their major difference being the slight shift of TM resonances toward higher frequencies, as can be seen in Fig.7(c). Figure 7(d) is a magnified view of the line-shape for a single TE resonant line centered at $\omega = 1.002068\omega_0$.

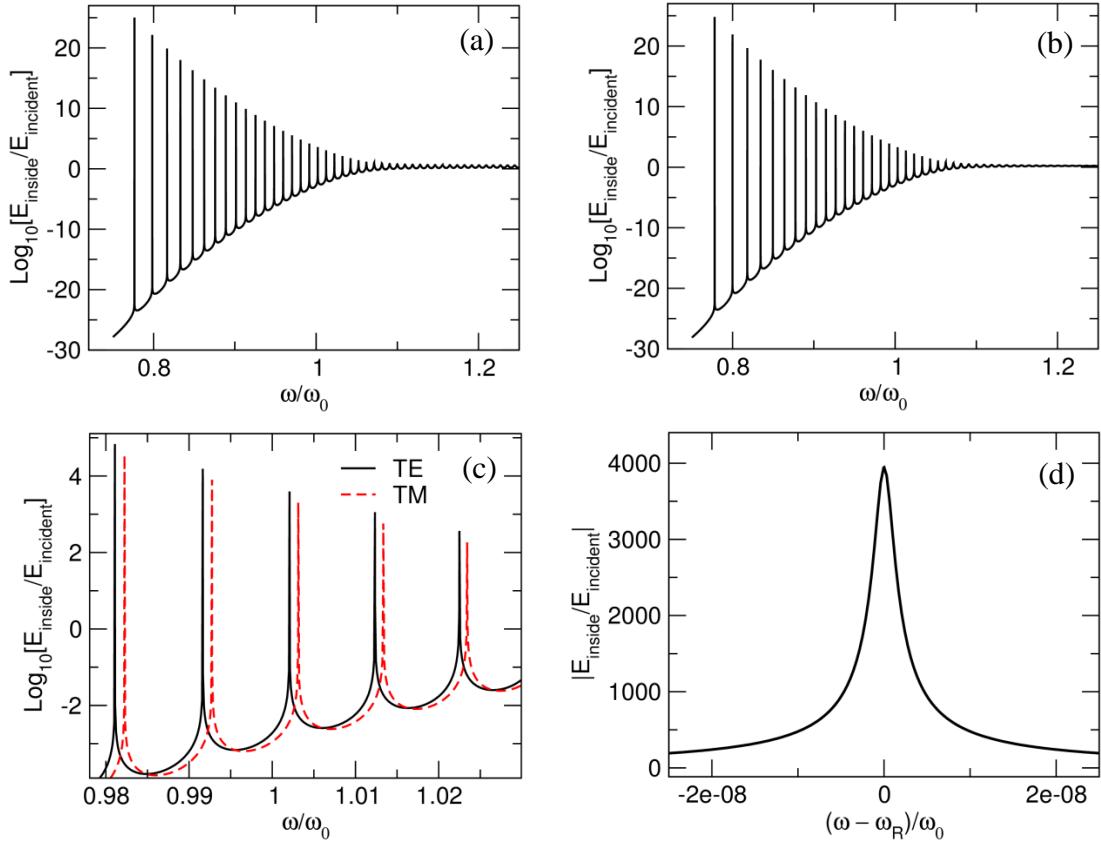

**Fig. 7**. Plots of the amplitude ratio of the $E$-field inside the dielectric sphere ($R = 50\lambda_0, n = 1.5$) to the incident $E$-field for the $\ell = 340$ spherical harmonic. The horizontal axis represents the excitation frequency $\omega$ normalized by $\omega_0 = 1.216 \times 10^{15}$ rad/sec. (a) TE mode. (b) TM mode. Note that the cutoff frequency for both modes is $\omega \cong 0.78\omega_0$, below which no resonances are excited. Above the cutoff, in between adjacent resonances, the field amplitude inside the cavity drops to exceedingly small values. The occurrence of extremely large resonance peaks in these plots is due to the assumed value of the refractive index $n$ being purely real. (c) Close-up view of the resonance lines of the glass ball for the $\ell = 340$ spherical harmonic, showing the TM resonances being slightly shifted away from the TE resonance lines. (d) Magnified view of an individual TE resonance line centered at $\omega_R = 1.002068\omega_0$.

To gain an appreciation for the effect of the mode number $\ell$ on the resonant behavior of our spherical cavity, we show in Fig.8 the computed ratio $E_{\text{inside}}/E_{\text{incident}}$ for $\ell = 10, 20$ and 25. It is observed that, with an increasing mode number $\ell$, the lowest accessible resonance moves to higher frequencies, and that the $Q$-factor associated with individual resonance lines tends to rise.

Finally, Fig.9 shows the computed $Q$-factors ($Q = |\omega_q'|/|\omega_q''|$) for a spherical cavity having $R = 50\lambda_0$, $n = n' + in''$, and $\ell = 340$. Setting $n' = 1.5$ allows a comparison between the results depicted in Fig.6, where $n'' = 0$, and those in Fig.9, which correspond to $n'' = 10^{-8}$ (blue), $10^{-7}$ (red), and $10^{-6}$ (black). These positive values of $n''$ account for the presence of small amounts of absorption within the spherical cavity. Compared to the case of $n'' = 0$, the resonance frequencies in Fig.9 have not changed by much, but the $Q$-factors of the various resonances are seen to have declined substantially. As expected, the greatest drop in the $Q$-factor is associated with the largest value of $n''$.

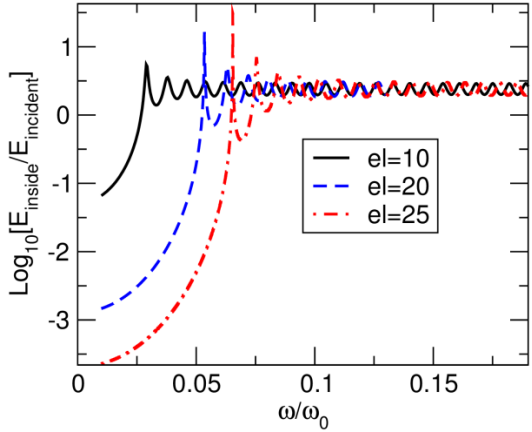 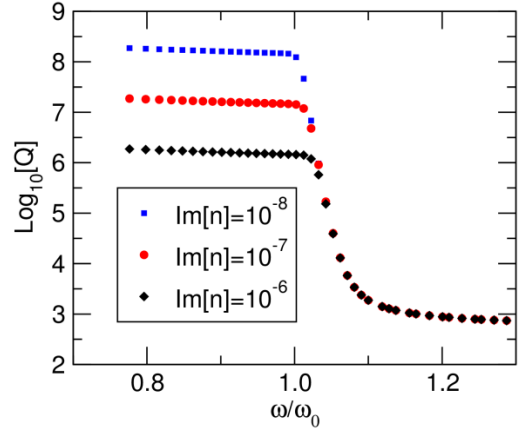

**Fig. 8**. Dependence on excitation frequency $\omega$ of the amplitude ratio of the $E$-field inside the glass sphere to the incident $E$-field for TE spherical harmonics having $\ell = 10$ (black), $\ell = 20$ (blue), and $\ell = 25$ (red).

**Fig. 9**. Similar to Fig. 6, except that the refractive index $n = n' + \mathrm{i}n''$ of the dielectric sphere is now allowed to have a small nonzero imaginary part, $n''$, representing absorption within the material.

**8. Leaky modes of a solid dielectric cylinder**. In a cylindrical coordinate system, within a linear, isotropic, homogeneous medium having permeability $\mu_0\mu(\omega)$, permittivity $\varepsilon_0\varepsilon(\omega)$, and refractive index $n(\omega) = \sqrt{\mu\varepsilon}$, Maxwell's equations have solutions of the form $\boldsymbol{E}(\boldsymbol{r},t) = \boldsymbol{E}(r)\exp[\mathrm{i}(m\varphi + k_z z - \omega t)]$ and $\boldsymbol{H}(\boldsymbol{r},t) = \boldsymbol{H}(r)\exp[\mathrm{i}(m\varphi + k_z z - \omega t)]$. Here the azimuthal mode-number $m$ could be a positive, zero, or negative integer, the *real-valued* $k_z$ is the propagation constant along the $z$-axis, and $\omega$ is the oscillation frequency [7]. The radial propagation constant will then be $k_r = \sqrt{[n(\omega)\omega/c]^2 - k_z^2}$, and the various field components will depend on the radial distance $r$ from the $z$-axis as follows:

$m = 0$ TE mode ($E_z = 0$):

$$\boldsymbol{E}(r) = E_0 J_1(k_r r)\widehat{\boldsymbol{\varphi}}. \tag{46}$$

$$\boldsymbol{H}(r) = -E_0[k_z J_1(k_r r)\widehat{\boldsymbol{r}} + \mathrm{i}k_r J_0(k_r r)\widehat{\boldsymbol{z}}]/[\mu_0\mu(\omega)\omega]. \tag{47}$$

$m \neq 0$ TE mode ($E_z = 0$):

$$\boldsymbol{E}(r) = E_0\big[\mathrm{i}(m/k_r r)J_{|m|}(k_r r)\widehat{\boldsymbol{r}} - \dot{J}_{|m|}(k_r r)\widehat{\boldsymbol{\varphi}}\big]. \tag{48}$$

$$\boldsymbol{H}(r) = E_0\big[k_z \dot{J}_{|m|}(k_r r)\widehat{\boldsymbol{r}} + \mathrm{i}(mk_z/k_r r)J_{|m|}(k_r r)\widehat{\boldsymbol{\varphi}} - \mathrm{i}k_r J_{|m|}(k_r r)\widehat{\boldsymbol{z}}\big]/[\mu_0\mu(\omega)\omega]. \tag{49}$$

$m = 0$ TM mode ($H_z = 0$):

$$\boldsymbol{E}(r) = H_0[k_z J_1(k_r r)\widehat{\boldsymbol{r}} + \mathrm{i}k_r J_0(k_r r)\widehat{\boldsymbol{z}}]/[\varepsilon_0\varepsilon(\omega)\omega]. \tag{50}$$

$$\boldsymbol{H}(r) = H_0 J_1(k_r r)\widehat{\boldsymbol{\varphi}}. \tag{51}$$

$m \neq 0$ TM mode ($H_z = 0$):

$$\boldsymbol{E}(r) = H_0\big[k_z \dot{J}_{|m|}(k_r r)\widehat{\boldsymbol{r}} + \mathrm{i}(mk_z/k_r r)J_{|m|}(k_r r)\widehat{\boldsymbol{\varphi}} - \mathrm{i}k_r J_{|m|}(k_r r)\widehat{\boldsymbol{z}}\big]/[\varepsilon_0\varepsilon(\omega)\omega]. \tag{52}$$

$$\boldsymbol{H}(r) = -H_0\big[\mathrm{i}(m/k_r r)J_{|m|}(k_r r)\widehat{\boldsymbol{r}} - \dot{J}_{|m|}(k_r r)\widehat{\boldsymbol{\varphi}}\big]. \tag{53}$$

In the above equations, $J_m(z)$ is a Bessel function of the first kind, integer-order $m$, and $\dot{J}_m(z)$ is its derivative with respect to $z$. One could also replace these with $Y_m(z)$ and $\dot{Y}_m(z)$, the Bessel function of the second kind and its derivative, or with $\mathcal{H}_m^{(1,2)}(z)$ and $\dot{\mathcal{H}}_m^{(1,2)}(z)$, the Hankel functions of type 1 and 2, and their corresponding derivatives. Useful identities include:

$$J_\nu(z) = (z/2)^\nu \sum_{k=0}^{\infty} \frac{(-1)^k (z/2)^{2k}}{k!\,\Gamma(\nu+k+1)}, \quad (|\arg(z)| < \pi). \quad \leftarrow \boxed{\text{G\&R 8.402}} \quad (54)$$

$$z\dot{J}_\nu(z) = \nu J_\nu(z) - z J_{\nu+1}(z). \quad \leftarrow \boxed{\text{G\&R 8.472-2}} \quad (55)$$

$$J_\nu(e^{im\pi} z) = e^{im\nu\pi} J_\nu(z). \quad \leftarrow \boxed{\text{G\&R 8.476-1}} \quad (56)$$

$$\boxed{\text{G\&R 8.451-1}} \rightarrow J_\nu(z) \sim \sqrt{2/(\pi z)}\, \cos(z - \tfrac{1}{2}\nu\pi - \tfrac{1}{4}\pi), \quad (|z| \gg 1 \text{ and } |\arg(z)| < \pi). \quad (57)$$

$$\boxed{\text{G\&R 8.451-3}} \rightarrow \mathcal{H}_\nu^{(1)}(z) \sim \sqrt{2/(\pi z)}\, \exp[\mathrm{i}(z - \tfrac{1}{2}\nu\pi - \tfrac{1}{4}\pi)], \quad (|z| \gg 1 \text{ and } |\arg(z)| < \pi). \quad (58)$$

$$z\dot{\mathcal{H}}_\nu^{(1)}(z) = \nu \mathcal{H}_\nu^{(1)}(z) - z \mathcal{H}_{\nu+1}^{(1)}(z). \quad \leftarrow \boxed{\text{G\&R 8.472-2}} \quad (59)$$

$$\mathcal{H}_\nu^{(1)}(e^{\mathrm{i}\pi} z) = -e^{-\mathrm{i}\nu\pi} \mathcal{H}_\nu^{(2)}(z). \quad \leftarrow \boxed{\text{G\&R 8.476-8}} \quad (60)$$

Consider now an infinitely-long, right-circular cylinder having radius $R$ and optical constants $\mu(\omega)$ and $\varepsilon(\omega)$, surrounded by free space. The radial propagation constants inside and outside the cylinder are denoted by $k_{r_0} = \sqrt{\mu\varepsilon(\omega/c)^2 - k_z^2}$ and $k_{r_1} = \sqrt{(\omega/c)^2 - k_z^2}$, respectively. For the $m = 0$ leaky TE mode, the boundary conditions at $r = R$ impose the following constraints:

$$E_0 J_1(k_{r_0} R) = E_1 \mathcal{H}_1^{(1)}(k_{r_1} R), \tag{61}$$

$$[E_0 k_{r_0}/\mu(\omega)] J_0(k_{r_0} R) = E_1 k_{r_1} \mathcal{H}_0^{(1)}(k_{r_1} R). \tag{62}$$

Therefore, for a leaky $m = 0$ TE mode to exist, the following characteristic equation must be satisfied:

$$F(\omega) = \mu(\omega) k_{r_1} \mathcal{H}_0^{(1)}(k_{r_1} R)\, J_1(k_{r_0} R) - k_{r_0} J_0(k_{r_0} R) \mathcal{H}_1^{(1)}(k_{r_1} R) = 0. \tag{63}$$

The corresponding equation for the leaky $m = 0$ TM modes is similar, with $\varepsilon(\omega)$ replacing $\mu(\omega)$. Of course, for a real-valued frequency $\omega$, if $|k_z|$ happens to be between $\omega/c$ and $n(\omega)\omega/c$, the EM field surrounding the dielectric cylinder will be evanescent, in which case the confined mode within the cylinder will *not* be leaky. Given that $K_m(z) = \tfrac{1}{2}\pi(\mathrm{i})^{m+1}\mathcal{H}_m^{(1)}(\mathrm{i}z)$, where $K_m(z)$ is a modified Bessel function of imaginary argument, one may rewrite Eq.(63) as follows:

$$\mu(\omega)\sqrt{k_z^2 - (\omega/c)^2}\, K_0\!\left(\sqrt{k_z^2 - (\omega/c)^2}\, R\right) J_1(k_{r_0} R) + k_{r_0} J_0(k_{r_0} R) K_1\!\left(\sqrt{k_z^2 - (\omega/c)^2}\, R\right) = 0. \tag{64}$$

When $\mu(\omega)$ and $\varepsilon(\omega)$ are real, Eq.(64) will have real-valued solutions for $\omega$, which represent the guided $m = 0$ modes of the cylinder. In general, however, the time-averaged Poynting vector associated with the field outside the cylinder will have a nonzero component along $\hat{r}$, indicating that Eq.(63) has solutions in the form of complex frequencies $\omega$ whose imaginary parts are negative.

Another special case occurs when $k_z = 0$, in which case the boundary conditions at $r = R$ impose the following constraints on TE modes:

$$E_0 \dot{J}_{|m|}(nR\omega/c) = E_1 \dot{\mathcal{H}}^{(1)}_{|m|}(R\omega/c), \tag{65}$$

$$[E_0 n(\omega)/\mu(\omega)] J_{|m|}(nR\omega/c) = E_1 \mathcal{H}^{(1)}_{|m|}(R\omega/c). \tag{66}$$

The corresponding characteristic equation will then be

$$F(\omega) = \mu(\omega)\dot{J}_{|m|}(nR\omega/c)\mathcal{H}^{(1)}_{|m|}(R\omega/c) - n(\omega)J_{|m|}(nR\omega/c)\dot{\mathcal{H}}^{(1)}_{|m|}(R\omega/c) = 0. \tag{67}$$

This equation is valid for positive, zero, and negative values of the azimuthal mode number $m$. It is also valid for TM modes provided that $\mu(\omega)$ is replaced with $\varepsilon(\omega)$. Note that, for $m = 0$, we have $\dot{J}_0(z) = -J_1(z)$ and $\dot{\mathcal{H}}^{(1)}_0(z) = -\mathcal{H}^{(1)}_1(z)$, confirming that Eq.(63) reduces to Eq.(67) when $k_z = 0$. In general, we expect the solutions of Eq.(67) to be complex frequencies $\omega$ having negative imaginary parts.

In the general case when $m \neq 0$ and $k_z \neq 0$, the boundary conditions at $r = R$ can be satisfied only for a superposition of TE and TM modes. Listed below are the continuity equations for $E_z$, $H_z$, $E_\varphi$, and $H_\varphi$. The continuity of $D_r$ is guaranteed by those of $H_z$ and $H_\varphi$, while the continuity of $B_r$ is guaranteed by those of $E_z$ and $E_\varphi$.[†]

$$E_z: \quad H_0 k_{r_0} J_{|m|}(k_{r_0}R) = H_1 \varepsilon(\omega) k_{r_1} \mathcal{H}^{(1)}_{|m|}(k_{r_1}R), \tag{68}$$

$$H_z: \quad E_0 k_{r_0} J_{|m|}(k_{r_0}R) = E_1 \mu(\omega) k_{r_1} \mathcal{H}^{(1)}_{|m|}(k_{r_1}R), \tag{69}$$

$$E_\varphi: \quad E_0 \dot{J}_{|m|}(k_{r_0}R) - i\left(\frac{mk_z}{\varepsilon_0 \varepsilon R\omega k_{r_0}}\right) H_0 J_{|m|}(k_{r_0}R) = E_1 \dot{\mathcal{H}}^{(1)}_{|m|}(k_{r_1}R) - i\left(\frac{mk_z}{\varepsilon_0 R\omega k_{r_1}}\right) H_1 \mathcal{H}^{(1)}_{|m|}(k_{r_1}R), \tag{70}$$

$$H_\varphi: \quad H_0 \dot{J}_{|m|}(k_{r_0}R) + i\left(\frac{mk_z}{\mu_0 \mu R\omega k_{r_0}}\right) E_0 J_{|m|}(k_{r_0}R) = H_1 \dot{\mathcal{H}}^{(1)}_{|m|}(k_{r_1}R) + i\left(\frac{mk_z}{\mu_0 R\omega k_{r_1}}\right) E_1 \mathcal{H}^{(1)}_{|m|}(k_{r_1}R). \tag{71}$$

The characteristic equation that ensures the existence of a non-trivial solution to the above equations is thus found to be

$$\left[\frac{\mu(\omega)}{k_{r_0}}\frac{\dot{J}_{|m|}(k_{r_0}R)}{J_{|m|}(k_{r_0}R)} - \frac{1}{k_{r_1}}\frac{\dot{\mathcal{H}}^{(1)}_{|m|}(k_{r_1}R)}{\mathcal{H}^{(1)}_{|m|}(k_{r_1}R)}\right] \times \left[\frac{\varepsilon(\omega)}{k_{r_0}}\frac{\dot{J}_{|m|}(k_{r_0}R)}{J_{|m|}(k_{r_0}R)} - \frac{1}{k_{r_1}}\frac{\dot{\mathcal{H}}^{(1)}_{|m|}(k_{r_1}R)}{\mathcal{H}^{(1)}_{|m|}(k_{r_1}R)}\right] = \left[\frac{mk_z}{R\omega/c}\left(\frac{1}{k_{r_0}^2} - \frac{1}{k_{r_1}^2}\right)\right]^2. \tag{72}$$

The values of $\omega$ that satisfy Eq.(72) are the leaky mode frequencies corresponding to the azimuthal mode number $m$ and the propagation constant $k_z$ along the $z$-axis. Note that, upon setting $m = 0$ in Eq.(72), the right-hand side of the equation vanishes. The two terms on the left-hand side will then be decoupled, representing the $m = 0$ TE and TM modes, respectively; this is consistent with the characteristic function given in Eq.(63). Similarly, setting $k_z = 0$ in Eq.(72) causes the right-hand side of the equation to vanish, thus, once again, decoupling the TE and TM modes, which are represented by the two terms on the left-hand side of the equation.

---

[†] $D_r: \; H_0 \dot{J}_{|m|}(k_{r_0}R) + i\left(\frac{\varepsilon_0 \varepsilon\, m\omega}{Rk_z k_{r_0}}\right) E_0 J_{|m|}(k_{r_0}R) = H_1 \dot{\mathcal{H}}^{(1)}_{|m|}(k_{r_1}R) + i\left(\frac{\varepsilon_0 m\omega}{Rk_z k_{r_1}}\right) E_1 \mathcal{H}^{(1)}_{|m|}(k_{r_1}R).$

$B_r: \; E_0 \dot{J}_{|m|}(k_{r_0}R) - i\left(\frac{\mu_0 \mu\, m\omega}{Rk_z k_{r_0}}\right) H_0 J_{|m|}(k_{r_0}R) = E_1 \dot{\mathcal{H}}^{(1)}_{|m|}(k_{r_1}R) - i\left(\frac{\mu_0 m\omega}{Rk_z k_{r_1}}\right) H_1 \mathcal{H}^{(1)}_{|m|}(k_{r_1}R).$

Given that, under the circumstances, $k_{r_0} = n(\omega)\omega/c$ and $k_{r_1} = \omega/c$, it is easy to verify that Eq.(72) is consistent with Eq.(67).

In the remainder of this section, our attention will be confined to TE modes with $k_z = 0$, for which the characteristic equation is given by Eq.(67). Our goal is to expand an initial field distribution residing within the cylinder at $t = t_0$, say, one of the mode profiles listed in Eqs.(46)-(53) having $k_z = 0$ and a real-valued oscillation frequency $\omega_0$, as a superposition of leaky modes whose complex frequencies $\omega = \omega_q$ are solutions of the characteristic equation $F(\omega) = 0$ given by Eq.(67). To this end, we form the function $G(\omega)$, as follows:

$$G(\omega) = \frac{\omega \exp(ik_{r_1}R)J_m(k_{r_0}r)/(k_{r_0}r)}{(\omega-\omega_0)F(\omega)}. \tag{73}$$

For odd (even) values of $m$, the numerator and the denominator of Eq.(73) are both even (odd) functions of the radial $k$-vector $k_{r_0}$. Consequently, $G(\omega)$ does *not* switch signs when $\omega$ crosses (i.e., goes from above to below) a branch-cut associated with $k_{r_0}(\omega)$. One must be careful here about $k_{r_1} = \sqrt{(\omega/c)^2 - k_z^2}$, which appears within the function $F(\omega)$; see Eqs.(63) and (72). The branch-cut associated with $k_{r_1}(\omega)$ is on the real-axis, spanning the interval $\text{Re}(\omega) \in [-ck_z, ck_z]$. When crossing this branch-cut, $k_{r_1}$ switches from a positive to a negative imaginary value, or vice-versa. The function $G(\omega)$ must remain insensitive to this sign-change as well. Unfortunately, we do not know how to handle this problem, and that is why we have limited the scope of the present discussion to the case of $k_z = 0$, for which the problem associated with the branch-cut residing on the real-axis disappears. (This problem is similar to that encountered in Sec.5 in the case of oblique incidence on a dielectric slab.)

In the limit when $|\omega| \to \infty$, the function $G(\omega)$ approaches zero (exponentially), thus ensuring the vanishing of $\oint G(\omega)d\omega$ over a large circle of radius $R_c$. Consequently, the residues of $G(\omega)$ evaluated at all its poles must add up to zero, that is,

$$\frac{\omega_0 \exp(i\omega_0 R/c)}{F(\omega_0)} \times \frac{J_m[n(\omega_0)\omega_0 r/c]}{n(\omega_0)\omega_0 r/c} + \sum_q \frac{\omega_q \exp(i\omega_q R/c)}{(\omega_q - \omega_0)F'(\omega_q)} \times \frac{J_m[n(\omega_q)\omega_q r/c]}{n(\omega_q)\omega_q r/c} = 0. \tag{74}$$

Rearranging the various terms of the above equation, we arrive at

$$J_m[n(\omega_0)\omega_0 r/c] = \sum_q \frac{n(\omega_0)F(\omega_0)\exp[i(\omega_q - \omega_0)R/c]}{(\omega_0 - \omega_q)n(\omega_q)F'(\omega_q)} \times J_m[n(\omega_q)\omega_q r/c]. \tag{75}$$

The above expansion may be used to represent the field components $E_r(r)$ of TE modes and $H_r(r)$ of TM modes; see Eqs.(48) and (53). To expand the field component $E_\varphi(r)$ of a TE mode or $H_\varphi(r)$ of a TM mode, the function $G(\omega)$ must be modified as follows:

$$G(\omega) = \frac{\omega \exp(ik_{r_1}R)J'_m(k_{r_0}r)}{(\omega-\omega_0)F(\omega)}. \tag{76}$$

In a similar vein, the appropriate forms of $G(\omega)$ for the remaining TE field components are

$$G(\omega) = \frac{\exp(ik_{r_1}R)J_m(k_{r_0}r)}{(\omega-\omega_0)\mu(\omega)F(\omega)}. \tag{77}$$

$$G(\omega) = \frac{\exp(ik_{r_1}R)J_m(k_{r_0}r)/(k_{r_0}r)}{(\omega-\omega_0)\mu(\omega)F(\omega)}. \tag{78}$$

$$G(\omega) = \frac{\exp(ik_{r_1}R)k_{r_0}J_m(k_{r_0}r)}{(\omega-\omega_0)\mu(\omega)F(\omega)}. \tag{79}$$

For TM modes, one must substitute $\varepsilon(\omega)$ for $\mu(\omega)$ in Eqs.(77)-(79).

**9. Numerical results for a cylindrical cavity.** Figure 10 provides a comparison between the resonances of a dielectric sphere ($R = 50\lambda_0, n = 1.5$) and those of a similar dielectric cylinder ($R = 50\lambda_0, L = \infty, n = 1.5, k_z = 0$). The excitation frequency $\omega$ is normalized by $\omega_0 = 1.216 \times 10^{15}$ rad/sec, which corresponds to the free-space wavelength $\lambda_0 = 1.55\ \mu m$. In Fig.10(a) the computed cavity $Q$-factors are plotted versus $\omega/\omega_0$ for the $\ell = 340$ TM spherical harmonic and the $m = 340$ TM cylindrical harmonic. Figure 10(b) compares the $E$-field amplitude ratio (i.e., $E$-field inside the cavity to the incident $E$-field) of the $\ell = 25$ TM spherical harmonic with that of the $m = 25$ TM cylindrical harmonic; this near agreement between the $E$-field ratios for the two cavities persists for larger values of $\ell$ and $m$. The resonances of the cylindrical rod are seen to be very similar to those of the glass sphere.

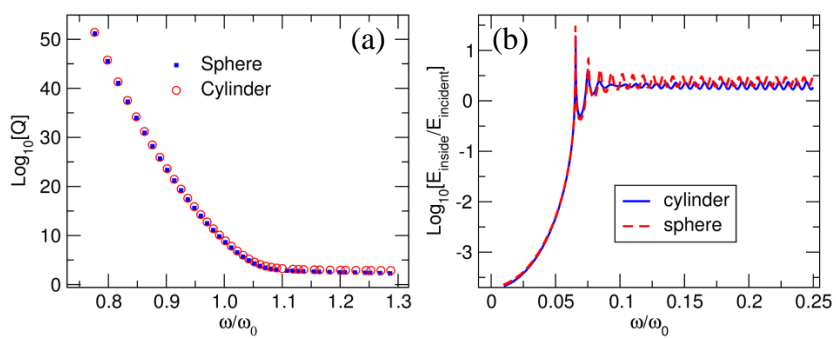

**Fig. 10**. A comparison between the resonances of a dielectric sphere ($R = 50\lambda_0, n = 1.5$) and a dielectric cylinder ($R = 50\lambda_0, L = \infty, n = 1.5, k_z = 0$). In (a) the cavity $Q$-factors are plotted versus $\omega/\omega_0$ for the $\ell = 340$ TM spherical and the $m = 340$ TM cylindrical harmonics. In (b) the ratio $E_{\text{inside}}/E_{\text{incident}}$ is plotted versus $\omega/\omega_0$ for the TM modes of the sphere ($\ell = 25$) and the cylinder ($m = 25$).

We thus observe that, in their general behavior, cylindrical cavities are quite similar to spherical cavities. This is not unexpected, considering that, for large $\ell$ and $m$, the EM field inside the sphere is more or less confined to a narrow band at the equator, and that the geometry of the equatorial region of a sphere is not too different from that of a cylinder. Of course, our calculations pertaining to the dielectric cylinder have been based on the assumption that the EM field is uniformly distributed along the cylinder axis, which is necessary if the results are to be compared with those for a spherical cavity at large $\ell$ values. Had we chosen, instead, to couple the light to the cylinder within a narrow strip (i.e., by illuminating a belt around the cylinder having a narrow spread along $z$), the light, once inside the cylinder, would have walked away from the strip due to diffraction effects. That would have caused a reduction in the $Q$-factor of the cylinder compared to that of a spherical cavity with a similar diameter at a large value of $\ell$.

**10. Concluding remarks.** Leaky modes of dielectric cavities contain a wealth of information about their resonant behavior, including the lifetimes associated with the light trapped inside the cavity immediately after the source of excitation is turned off. We have proved the completeness of these leaky modes under special circumstances, although completeness under more general conditions remains to be demonstrated. Our completeness proof rigorously accounts for realistic dispersion effects, including absorption losses and the existence of branch-cuts associated with the Lorentz oscillator model. Our numerical results have intimated the close connection between resonant behavior and the leaky eigen-modes of dielectric slabs, spheres, and cylinders. The fact that spherical harmonics with large $\ell$ values, and also cylindrical harmonics with large $m$ values, are associated with high-$Q$ resonances hints at the importance of EM angular momentum in relation to the long lifetimes of the modes trapped inside these cavities. In other words, there appears to be a connection between the strength of the circular motion of EM energy inside a cavity and the time it takes for this energy to leak out. These connections will be explored in a forthcoming publication.

## Appendix

We show that $F(\omega)$ of Eq.(37) approaches a constant when $\omega \to 0$. In the limit $z \to 0$, we have

$$J_\nu(z) \to \frac{(z/2)^\nu}{\Gamma(\nu+1)}. \quad \leftarrow \text{G\&R 8.440} \quad (A1)$$

$$Y_\nu(z) \to \frac{(z/2)^\nu}{\tan(\nu\pi)\Gamma(1+\nu)} - \frac{(z/2)^{-\nu}}{\sin(\nu\pi)\Gamma(1-\nu)}; \quad (\nu \neq \text{an integer}). \quad \leftarrow \text{G\&R 8.443} \quad (A2)$$

Therefore, when $\omega \to 0$, we will have

$$F(\omega) = nk_0 R \mathcal{H}^{(1)}_{\ell+\frac{1}{2}}(k_0 R) J_{\ell+3/2}(nk_0 R) + [(\mu-1)(\ell+1)\mathcal{H}^{(1)}_{\ell+\frac{1}{2}}(k_0 R) - \mu k_0 R \mathcal{H}^{(1)}_{\ell+3/2}(k_0 R)] J_{\ell+\frac{1}{2}}(nk_0 R) \quad \leftarrow k_0 = \omega/c$$

$$= nk_0 R J_{\ell+\frac{1}{2}}(k_0 R) \overset{0}{\cancel{J_{\ell+3/2}(nk_0 R)}} + [(\mu-1)(\ell+1)J_{\ell+\frac{1}{2}}(k_0 R) \overset{0}{\cancel{-\mu k_0 R J_{\ell+3/2}(k_0 R)}}] J_{\ell+\frac{1}{2}}(nk_0 R)$$

$$+ ink_0 R Y_{\ell+\frac{1}{2}}(k_0 R) J_{\ell+3/2}(nk_0 R) + i[(\mu-1)(\ell+1)Y_{\ell+\frac{1}{2}}(k_0 R) - \mu k_0 R Y_{\ell+3/2}(k_0 R)] J_{\ell+\frac{1}{2}}(nk_0 R)$$

$$\to \frac{i(-1)^{\ell+1} nk_0 R}{\Gamma(\frac{1}{2}-\ell)\Gamma(\ell+5/2)} (\frac{1}{2}k_0 R)^{-(\ell+\frac{1}{2})} (\frac{1}{2}nk_0 R)^{\ell+3/2}$$

$$+ \frac{i}{\Gamma(\ell+3/2)} \left[ \frac{(-1)^{\ell+1}(\mu-1)(\ell+1)}{\Gamma(\frac{1}{2}-\ell)} (\frac{1}{2}k_0 R)^{-(\ell+\frac{1}{2})} - \frac{(-1)^\ell \mu k_0 R}{\Gamma(-\frac{1}{2}-\ell)} (\frac{1}{2}k_0 R)^{-(\ell+3/2)} \right] (\frac{1}{2}nk_0 R)^{\ell+\frac{1}{2}}$$

$$\to \frac{i(-1)^{\ell+1} n^{\ell+\frac{1}{2}}}{\Gamma(\ell+3/2)} \left[ \frac{(\mu-1)(\ell+1)}{\Gamma(\frac{1}{2}-\ell)} + \frac{2\mu}{\Gamma(-\frac{1}{2}-\ell)} \right]. \quad \leftarrow n(0) = \sqrt{\mu(0)\varepsilon(0)} \quad (A3)$$

Consequently, $F(\omega)$ has no poles at $\omega = 0$, which indicates that, in the vicinity of $\omega = 0$, the function $G(\omega)$ is not singular.

**Acknowledgement**. This work has been supported in part by the AFOSR grant No. FA9550-13-1-0228.

---

[†] A general representation of Bessel functions of the first kind, order $\nu$, is $J_\nu(z) = (z/2)^\nu \sum_{k=0}^\infty \frac{(-1)^k (z/2)^{2k}}{k!\,\Gamma(\nu+k+1)}$ (G&R 8.440). Considering that, for spherical harmonics, $\nu = \ell + \frac{1}{2} \geq 3/2$, it is seen that $J_{\ell+\frac{1}{2}}(z)/z \to 0$ when $z \to 0$.